\documentclass[11pt]{article}
\usepackage{latexsym}
\usepackage{amsfonts}
\usepackage{amsmath}
\usepackage{amsthm}
\usepackage{graphicx}
\usepackage{epic}
\def\a{\alpha}
\def\b{\beta}
\def\g{\gamma}
\def\d{\delta}

\def\l{\lambda}

\def\m{\mu}
\def\n{\nu}
\def\r{\rho}

\def\p{\pi}

\def\cA{{\mathcal A}}

\def\be{\begin{equation}}
\def\ee{\end{equation}}
\def\beq{\begin{eqnarray}}
\def\eeq{\end{eqnarray}}
\def\nn{\nonumber}

\def\ce{{\mathcal E}}
\def\cf{{\mathcal F}}

\def\ch{{\mathcal H}}

\def\ck{{\mathcal K}}

\def\cm{{\mathcal M}}
\def\cn{{\mathcal N}}

\def\cv{{\mathcal V}}

\def\RR{{\mathbb{R}}}

\def\ZZ{{\mathbb{Z}}}

\newcommand{\ft}[2]{{\textstyle {\frac{#1}{#2}} }}

\newcommand{\E}{E_{10}}

\newcommand{\Ref}[1]{(\ref{#1})}
\newcommand{\non}{\nonumber\\}

\newcommand{\bqn}{\begin{eqnarray}}\newcommand{\eqn}{\end{eqnarray}}

\begin{document}

\begin{titlepage}

\begin{centering}

{\huge {\bf Poincar\'e, Relativity, Billiards and Symmetry}}

\vspace{1.5cm}

\centerline{\large{Thibault Damour}} \vspace{.7cm}
\centerline{ {\it Institut
des Hautes Etudes Scientifiques}, } \centerline{{\it 35, Route de
Chartres,  F-91440 Bures-sur-Yvette, France} }

\vspace{5cm} \hrule

\begin{abstract}
 This review is made of two parts which are related to Poincar\'e  in different ways.
The first part reviews the work of Poincar\'e on the Theory of (Special) Relativity.
One emphasizes both the remarkable achievements of Poincar\'e, and the fact that he
never came close to what is the essential conceptual achievement of Einstein: changing
the concept of time. The second part reviews a topic which probably would have appealed to Poincar\'e
because it involves several mathematical structures he worked on: chaotic dynamics, discrete
reflection groups, and Lobachevskii space. This topic is the hidden role of Kac-Moody
algebras in the billiard description of the asymptotic behaviour of certain Einstein-matter systems
near a cosmological singularity. Of particular interest are the Einstein-matter systems arising in the
low-energy limit of superstring theory. These systems seem to exhibit the highest-rank hyperbolic
Kac-Moody algebras, and notably $\E$, as hidden symmetries.

\end{abstract}
\hrule

\vspace{3mm}

\end{centering}

\vfill
\end{titlepage}

\tableofcontents

\section{POINCAR\'E and RELATIVITY}

\subsection{Some significant biographical dates}

Starting in 1886, Poincar\'e holds the chair of ``Physique math\'ematique et calcul des probabilit\'es" at the Sorbonne. His 1888 lectures are about Maxwell's theories and the electromagnetic theory of light. In 1889, 1892 and 1899, he lectures on the works by Helmholtz, Hertz, Larmor and, especially, Lorentz. His lectures give him the matter of four books in which he expounds all the different theories. This pedagogical work makes him fully aware of the state of the art in modern
electromagnetic theories, and establishes him as a renowned expert in mathematical physics. He exchanges a correspondence with Lorentz (who keeps him informed of his work),
and starts writing some research papers on electromagnetic theory.

In 1893, Poincar\' e becomes a member of the ``Bureau des Longitudes". It is the time where one starts to use telegraphic signals for synchronizing clocks (Br\'eguet, 1857; Le Verrier, 1875,...) and for measuring longitudes (see \cite{Galison}).

In 1900,  the ``Congr\`es International de Physique" takes place in Paris, and Poincar\'e gives an invited
review talk on the ``Relations entre la physique exp\'erimentale et la physique math\'ematique". Four years later, on September 24, 1904, during the ``Congr\`es international des Arts et des Sciences" of Saint-Louis, Missouri (USA),
Poincar\'e gives an invited review talk on  ``L'\'etat actuel et l'avenir de la physique math\'ematique".

In 1902, Poincar\'e publishes his popular book ``La Science et l'hypoth\`ese"; more than 16000 copies of this book have been sold. In 1905, he publishes his second popular book: ``La valeur de la Science". Two other
popular books shall come out later:  ``Science et M\'ethode", in 1908, and ``Derni\`eres Pens\'ees",
posthumously,  in 1913.

\subsection{Selected citations and contributions of Poincar\'e to Relativity}

In 1898, in a paper entitled ``La Mesure du temps" \cite{P98}, he writes: ``Nous n'avons pas l'intuition directe de la simultan\'eit\'e, pas plus que celle de l'\'egalit\'e de deux dur\'ees."
He discusses in detail the fact that, in order to define time and simultaneity, one must admit, as postulates, some ``rules", e.g. that the velocity of light is constant and isotropic, and then that one must correct for the non-zero
 transmission times when using telegraphic signals to synchronize faraway clocks.

In the paper ``La th\'eorie de Lorentz et le principe de r\'eaction" \cite{P00}, written in 1900 at the occasion of the 25th anniversary of Lorentz's thesis, Poincar\'e discusses (as emphasized in \cite{Darrigol}) the effect of an overall
translation, at some speed $v$, on the synchronization of clocks by the exchange of electromagnetic signals.
More precisely, he works only to {\it first order} in $v$, and notes that, if moving observers synchronize their watches by exchanging optical signals, and if they correct these signals by the transmission time under the (incorrect) assumption that the signals travel at the same speed in both directions, their watches will indicate not the ``real time'', but  the ``apparent time", say
\be \tau = t - \frac{vx}{c^2} + O(v^2).
\label{1} \ee
His main point is that the ``apparent time'' $\tau$ coincides with the formal mathematical variable
$t' \equiv t - \frac{vx}{c^2} + O(v^2)$ introduced
by Lorentz in 1895 under the name of ``local time'' (and used by him to show the invariance of Maxwell theory
under uniform translations, to first order in $v$).

In the book ``La Science et l'hypoth\`ese", dated from 1902\footnote{This seems to be the only work of Henri Poincar\'e read by Einstein before 1905.}, Poincar\'e writes suggestive sentences such as:
\begin{itemize}

\item ``Il n'y a pas d'espace absolu et nous ne concevons que des mouvements relatifs."

\item ``Il n'y a pas de temps absolu."

\item ``Nous n'avons pas l'intuition directe de la simultan\'eit\'e de deux \'ev\'ene-\\ments."


\item `` $\cdots$ je ne crois pas, malgr\'e Lorentz, que des observations plus pr\'ecises puissent jamais mettre en \'evidence autre chose que les d\'eplacements relatifs des corps mat\'eriels." [In fact, this is reprinted from his 1900 talk at
the Congr\`es International de Physique.]
\end{itemize}

Poincar\'e recalls that when experiments  testing effects of first order in $v$
 all came out negative, Lorentz found a general explanation at this order $O(v^1)$. When experiments testing the order $v^2$ again gave negative results (Michelson-Morley 1887), one found an ad hoc hypothesis (namely the Lorentz-FitzGerald contraction). For Poincar\'e, this is clearly unsatisfactory:

  ``Non, il faut trouver une m\^eme explication pour les uns et pour les autres et alors tout nous porte \`a penser que  cette explication vaudra \'egalement pour tous les termes d'ordre sup\'erieurs et que la destruction mutuelle de ces termes sera rigoureuse et absolue".

On Mai 27, 1904, Lorentz publishes his crucial paper: ``Electromagnetic phenomena in a system moving with any velocity smaller than that of light".
This paper contains the full ``Lorentz transformations'' linking variables associated to a moving
  frame, say $(t',x',y',z')$, to the ``true, absolute'' time and space coordinates $(t,x,y,z)$. [Lorentz considers $(t',x',y',z')$
only as convenient mathematical variables.]

On September 24, 1904, at the Saint-Louis ``Congr\`es International", and later in his 1905 popular book ``La Valeur de la Science", Poincar\'e mentions the ``Principe de Relativit\'e" among a list of the  basic principles of physics.

Then he mentions that Lorentz's ``local time" is the (apparent) time indicated by moving clocks (say, $A$ and $B$) when
they are synchronized by exchanging light signals and by (`wrongly' but conventionally)
assuming the isotropy of the speed of
light in the moving frame, {\it i.e.} the equality between the transmission times during the two exchanges
${A\to B}$ and ${B\to A}$. [However, he does not write down any equations, so that it is not clear
whether he is alluding to his previous {\it first order in $v$} result, (\ref{1}), or to an all order result (see below)].

He also mentions the existence of a ``m\'ecanique nouvelle" where inertia goes to $\infty$ as $v\to c$, and therefore, $c$ is a limiting speed. [Note, however, that this was a feature common to all the current electron dynamics: Lorentz's, Abraham's, etc.]

On June 5, 1905\footnote{Notice that Einstein's paper on Relativity was received by the {\it Annalen der Physik}
 on June 30, 1905.}, Poincar\'e
submits to the Comptes Rendus of the Acad\'emie des Sciences the short Note:
 ``Sur la dynamique de l'\'electron" \cite{P05}. This is followed by a more detailed article \cite{P06}
 (received on July 23, 1905). In these papers:
\begin{itemize}
\item he admits ``sans restriction" le ``Postulat de Relativit\'e" and explores its consequences
\item he modifies and completes Lorentz's 1904 paper (by giving the correct transformation laws
for the electromagnetic field, and source, quantities).
\item he explains ``dynamically" why each ``electron" undergoes the Lorentz contraction by introducing a (negative) internal pressure holding the electron against its electric self-repulsion
\item he proves the group structure of ``Lorentz transformations" [a name that he introduces in these papers]
\item he proves the invariance of
\be \Delta s^2 = (\Delta x)^2 + (\Delta y)^2 + (\Delta z)^2 - c^2\,(\Delta t)^2
\ee
\item he introduces $\ell = ict$ and makes the identification of Lorentz transformations
 with rotations in a 4-dimensional euclidean space
\item he proves the addition law for velocity parameters
\be w=\frac{u+v}{1+\frac{uv}{c^2}}
\ee
\item he discusses the invariants and covariants of the Lorentz group, e.g.
\be \vec E^{\,2} - \vec B^{\,2} = \mbox{invariant}\quad\quad,\quad\quad x^\mu \sim J^\mu\ee
\item he demands that the Principle of Relativity apply to gravitation
\item he discusses possible relativistic laws of gravity (of an action-at-a-distance type, {\it i.e.} without
assuming any explicit field content)
\item he mentions that relativistic ``retarded" gravitational interactions propagating with velocity $c_g=c$ are in agreement with existing observational  limits on $c_g$ (due to Laplace) because they all differ from Newton's
law only at the order  $O(v^2/c^2)$
\item he speaks of ``ondes gravifiques" both in the sense of retarded interaction and of emission of radiation
\item he concludes about the necessity of a more detailed discussion of $O(v^2/c^2)$ deviations from
Newtonian gravity.
\end{itemize}

In 1906-1907, Poincar\'e's Sorbonne lectures \cite{P53} (published in 1953 !) are about ``Les limites de la loi de Newton":
\begin{itemize}
\item In them Poincar\'e sets a limit $\vert \frac{\Delta a}{a}\vert \leq 2\times 10^{-8}$ on the ratio between the
gravitational mass and the inertial mass, $m_{gravit}/m_{inertia}$, by updating Laplace's work on the
polarization of the  Earth-Moon orbit by the Sun. [This effect has been rediscovered, in a different context,
by Nordtvedt in 1968.]
\item he discusses observational consequences of (among many others possible modifications of Newton's law) some selected ``relativistic" laws of gravity and shows that their main observational effect is an additional
advance of the perihelion of Mercury: e.g. he mentions
 that an electromagnetic-like gravitational law (``spin-1 exchange") yields an additional perihelion
 advance of  $7"$ per century\footnote{This had been already derived by Lorentz in 1900. Poincar\'e knew, however,
 at this stage that Lorentz's electromagnetic-like gravitational law was just one possibility among many ``relativistic laws''.}, instead of an observed value that he quotes as $\sim 38"$\footnote{It is amusing to
 speculate about what would have happened if Poincar\'e had used the better value (obtained by the american astronomer
 Simon Newcomb at the end of the 19th century) of $\sim 43"$, and had noticed that the various relativistic results
 he was deriving were all {\it integer} submultiples of the observed value ($1/6$ in the case of spin-1 exchange). }
 at the time)
\item he works out the synchronization of moving clocks, by the method he had already mentioned in 1900-1904, to all orders in $v/c$ and {\it seems} to conclude (see, however, below) that the result is exactly the (all orders) ``local time'' $t'$ introduced
by Lorentz in 1904
 \be t' = \frac{t-vx/c^2}{\sqrt{1-v^2/c^2}} = \sqrt{1-v^2/c^2} \left[ t - \frac{v}{c^2-v^2}(x-vt)  \right]
\label{16}\ee
\item he determines how one must modify the fundamental law of dynamics $ F=ma$ so that the principle of relativity holds
\end{itemize}

In 1908, in a paper entitled ``La dynamique de l'\'electron" \cite{P08} (which is essentially reprinted in his 1908 book  ``Science et M\'ethode"), Poincar\'e speaks more about gravitational waves. More precisely, he mentions that the main observable effect of the ``onde gravifique" emitted at infinity by an orbiting system (``onde d'acc\'el\'eration") will be, because of radiation reaction in the source, a secular acceleration of the orbital frequency, {\it i.e.} a negative value of the
orbital period $P$: $\dot P<0$. This observable effect  is exactly what has been measured in binary pulsars, such as PRS 1913+16,
which provided the first direct proofs that gravity propagates with the velocity of light (see, e.g., \cite{PDG}).

\subsection{Assessment of Poincar\'e's contributions to Relativity.}

The above list of statements and results is certainly an impressive list of achievements! Some people (have) claim(ed) that Poincar\'e should share, with Einstein, the credit for discovering the ``Theory of Relativity". When
discussing this matter one should carefully distinguish various aspects of Poincar\'e's contributions.

Technically, it is true that Poincar\'e made important new contributions related to what is now called
Special Relativity. Notably, the action of the Lorentz group on electromagnetic variables $A_{\mu}, F_{\mu \nu}, J_{\mu}$;
the group structure of Lorentz transformations; the invariance of the spacetime interval
$ \Delta s^2 = (\Delta x)^2 + (\Delta y)^2 + (\Delta z)^2 - c^2\,(\Delta t)^2$; and the proposal to consider  the ``Principle
of Relativity" as a general  principle of physics applying, for instance, to gravitation and thereby restricting possible
relativistic generalizations of Newton's law.  For all those technical achievements, it would be quite reasonable to imagine that, if a Nobel prize
had been given for Special Relativity before the death of Poincar\'e in 1912, the prize could have been shared
between Einstein, Lorentz and Poincar\'e.

However, at the {\it conceptual} level, it seems to me
(in spite of contrary claims in the literature) that Einstein was the first one to make what is the crucial step of
Special Relativity, namely proposing a revolutionary\footnote{Let us note that Max Planck was the first
scientist who understood the revolutionary nature of the new einsteinian conceptual setup. In 1910 he wrote that
``This principle [of Relativity] has brought about a revolution in our physical picture of the world, which, in
extent and depth, can only be compared to that produced by the introduction of the Copernican world system.''
He also wrote that ``In boldness it [Special Relativity] probably surpasses anything so far achieved in
speculative natural science, and indeed in philosophical cognition theory; non-Euclidean
 geometry is child's play in comparison."\cite{Folsing}.}
change in the concept ot {\it time}. Moreover, as we shall see below, Poincar\'e resisted till his death
such a change in the concept of time.

This conceptual revolution in the notion of time is encapsulated in the ``twin paradox'', {\it i.e.}
in time dilation effects, much more than in any change of synchronization conventions. Indeed, it was the
 idea that the variable $t'$ was ``time, pure and simple'' which led Einstein, for the first time, to think
and predict that, independently of any synchronization convention, a clock  moving away and then coming back will not
mark the same time when it reconvenes with its ``sister clock'' that remained in inertial motion. It is true that
Poincar\'e's discussion of synchronization in a moving frame seems close to Einstein's synchronization process,
but, when looking more carefully at what Poincar\'e actually wrote, one finds that there is a world of difference
between the two.

First, let us mention that all the papers of Poincar\'e dealing with clock synchronization and
published before Einstein's 1905 work on Special Relativity either dealt only with $O(v^1)$ effects (at which
order there are no time dilation effects), or contained no explicit formulas [as in his Saint-Louis, September 1904 lecture].
The only explicit work of Poincar\'e on clock synchronization which keeps all orders in $v/c$ is {\it posterior }
to Einstein's 1905 paper on Relativity. It is contained in
his 1906-7 Sorbonne lectures (published only in 1953 \cite{P53}) or in his 1908 paper \cite{P08}.

\begin{itemize}
 \item When looking in detail
at the results actually derived by Poincar\'e, both in \cite{P53} and (consistently) in \cite{P08}, one finds that
Poincar\'e actually derives the following expression for what he calls the ``apparent time'' (``temps apparent'')
marked moving clocks in the way he advocates:

\be
 \tau_{\mbox{(Poincar\'e)}} = t - \frac{v}{c^2-v^2}(x-vt) \equiv \frac{1}{\sqrt{1-v^2/c^2}} t'
\label{17} \ee
  with $t'$ given by (\ref{16}). In other words, $t'$ is the result of Einstein for the ``time'' in the moving frame
 (previously introduced by Lorentz as a mathematically auxiliary ``local time'' variable).

 The crucial point is that
 Poincar\'e's synchronized time $\tau$,(\ref{17}), {\it differs from Einstein's}  ``time, pure and simple'', $t'$, in the moving frame
{\it precisely by the time-dilation factor}  $\gamma =1/{ \sqrt{1-v^2/c^2} }$. In other words, though moving clocks marking
 the Poincar\'e time $\tau$ are desynchronized among themselves with respect to the absolute time (``temps vrai'') $t$, (because of the $(x-vt)$-term in (\ref{17})), they
 do beat the same  ``absolute time'' as a clock at rest, $ d \tau = d t$, and do not exhibit any ``twin paradox''.
  [This is consistent with Poincar\'e's
 statements, both in his 1904 Saint-Louis lecture, and in ``La Valeur de la Science'', that, among two moving clocks
 ``l'une d'elles retardera sur l'autre''. The context shows that he does not speak of Einstein's time retardation
 effects linked to the factor  $\gamma =1/{\sqrt{1-v^2/c^2} }$ but of the constant offset
 $\propto  - \frac{v}{c^2-v^2} \Delta (x-vt)$ between the indications of two clocks moving with the same,
 uniform velocity $v$.] It is true that in a subsequent paragraph of \cite{P53} Poincar\'e seems to identify the result
  $\tau$,(\ref{17}), of his explicit calculation with the full Lorentz ``local time'' $t'$. However, it seems clear to me
 that, in doing that, Poincar\'e has missed ``thinking'' the crucial einsteinian revolutionary step. {\it Mathematically},
 Poincar\'e knew that the variable with good properties was $t'$ (and this is the ``time variable'' he uses in
 his important technical papers of 1905 \cite{P05,P06}), but {\it physically} he never thought, nor proposed, that
 a moving clock will mark the time $t'$ (and will therefore exhibit a ``twin paradox'').

 Additional evidence for this  limitation of the horizon of thought of Poincar\'e comes from other statements of his:

\item Poincar\'e  always distinguishes ``le temps vrai", $t$,  from ``le temps apparent", $\tau$ or $t'$, and,
similarly, he always thinks in terms of ``absolute space"
\item Poincar\'e kept asking for a deeper (dynamical ?) reason behind the ``relativity postulate", and the appearance of the velocity of light ($c_{light}$) in possible relativistic laws of gravity
\item Poincar\'e had no firm theoretical a priori conviction in the ``relativity principle"; e.g.
\end{itemize}
- In his 1908 paper \cite{P08} (and in his 1908 book on ``Science et M\'ethode"\footnote{Though he added
a last minute footnote stating that the more recent experiments of Bucherer agreed with  relativistic
dynamics.}), commenting Kaufmann's early experiments (that did not seem to
quite agree with relativistic dynamics), he expresses doubts about the exact validity
of the relativity principle: He writes that the latest experiments of Kaufmann

``{\it ont donn\'e raison \`a la th\'eorie d'Abraham.}
Le Principe  de Relativit\'e n'aurait donc pas la valeur rigoureuse qu'on \'etait tent\'e de lui attribuer; $\cdots$''

By contrast, Einstein, commenting in 1907 the {\it same} experimental results, states that the agreement
with relativity is rather good  in first approximation, and that the deviations are probably due
to systematic errors. Indeed, Einstein writes that it is a priori more probable that relativistic dynamics be correct, rather
than Abraham's dynamics, because the former is based on a general principle having
wide ranging consequences for physics as a whole.

- in his 1906-1907 lectures, Poincar\'e concludes that the most probable explanation
for the anomaly in Mercury's perihelion advance is the existence of an infra-mercurial ring of matter.
He has not  enough trust in any of the possible relativistic gravitational theories he had introduced in 1905
to propose their  $ v^2/c^2$ effects as a likely explanation for it. [As we said above,  had he noticed that they
all gave an {\it integer} submultiple of the observed anomaly, he might have suspected that one of them
might give the correct explanation.]

- in 1904 (Saint-Louis), 1905 (``La valeur de la science''), and in 1908 ({\it i.e.} several years after Einstein's famous September 1905 paper on $E=mc^2$), Poincar\'e speaks of the recoil in reaction to the emission of electromagnetic waves and says : ``Ce que nous avons envoy\'e au loin, ce n'est plus un projectile mat\'eriel, c'est de l'\'energie et l'\'energie n'a pas de masse", {\it i.e.} ``energy has no mass'' 	! (``Science et m\'ethode'', livre III,
chapter II, 1908)

- in 1912, a few months before his death, Poincar\'e writes \cite{P13} some sentences that have been
quoted as evidence for Poincar\'e's role in conceptualizing (or at least accepting the Einstein-Minkowski)
 spacetime as a physical structure. For instance,
he writes: ``Tout se passe comme si le temps \'etait une quatri\`eme dimension de l'espace; et comme si
l'espace \`a quatre dimensions r\'esultant de la combinaison de l'espace ordinaire et du temps pouvait tourner
non seulement autour d'un axe de l'espace ordinaire [$\cdots$] mais autour d'un axe quelconque. [$\cdots$] dans
la nouvelle conception l'espace et le temps ne sont plus deux entit\'es enti\`erement distinctes et que l'on puisse
envisager s\'epar\'ement, mais deux parties d'un m\^eme tout [$\cdots$] qui sont comme \'etroitement
enla\c{c}\'ees $\cdots$ ".
However, if one reads the full text, one realizes that Poincar\'e explains here a conception proposed by
``some physicists'', and that he is not at all ready to accept this new conception (or ``convention'').
Indeed, he ends his text by writing:

``Aujourd'hui certains physiciens veulent adopter une convention nouvelle [$\cdots$] ceux qui ne sont pas de cet avis peuvent l\'egitimement conserver l'ancienne pour ne pas troubler leurs vieilles habitudes. Je crois, entre nous, que c'est ce qu'ils feront encore longtemps."

This last sentence, which constitutes the last words written by Poincar\'e on Special Relativity,  shows
clearly that Poincar\'e never believed in the physical relevance of the conceptual revolution brought
by Einstein in the concept of time (and extended by Minkowski to a revolutionary view
of the physical meaning of spacetime).

\subsection{Poincar\'e on Einstein, concerning Relativity}
\begin{itemize}
\item Poincar\'e never mentioned Einstein's work on relativity (neither in his papers or books,
nor, as far as I know, in his letters). Poincar\'e seemed to be unaware of Einstein's work during
the years 1906-1909. His attention was probably brought to the work of Einstein and Minkowski
only in the spring of 1909. [In April 1909 Poincar\'e gave some lectures in G\"ottingen, notably
on ``la m\'ecanique nouvelle'', without mentioning the names of Einstein or Minkowski.]

\item Maybe he thought that 

- technically, there was nothing new in Einstein's work on Relativity

- conceptually, Einstein was  ``cheating"\footnote{I thank David Gross for a useful discussion
on this point.} because he simply assumed (kinematically) what had to be proven (dynamically) [as Lorentz thought, and as Poincar\'e's electron-pressure model contributed to proving]

- Einstein's real contribution remained physically obscure to him, because Poincar\'e always thought that ``apparent time" should be different from ``real time" (while Einstein summarized his main contribution as being
 ``the realization that an auxiliary term introduced by H. A. Lorentz and called by him `local time'
could be defined as `time pure and simple'.'').

\item As a final vista on the conceptual difference between Poincar\'e and Einstein, let us mention the
following revealing anecdote. Poincar\'e and Einstein met only once, at the Solvay meeting of 1911. Maurice de Broglie
 (who was one of the secretaries of this first Solvay meeting) wrote (in 1954 \cite{dB54})
the following:

``Je me rappelle qu'un jour \`a Bruxelles, comme Einstein exposait ses id\'ees [sur la ``m\'ecanique nouvelle''
c'est-\`a-dire relativiste],  Poincar\'e lui demanda:
`  Quelle m\'ecanique adoptez-vous dans vos raisonnements ?'  Einstein lui
r\'epondit :`Aucune m\'ecanique', ce qui parut surprendre son interlocuteur."

This conversation on  ``relativistic mechanics''
 (which contrasts the ``dynamical approach'' of Poincar\'e to the ``kinematical'' one
of Einstein) is not reported in the official proceedings of the 1911 Solvay meeting which concerned
the (old) theory of quanta.

\end{itemize}

\section{RELATIVITY, BILLIARDS and SYMMETRY}

\subsection{Introduction and overview}

A remarkable connection between the asymptotic behavior
of certain Einstein-matter systems near a cosmological singularity  and
billiard motions in the Weyl chambers of some corresponding Lorentzian
Kac--Moody algebras was uncovered in a series of works \cite{DH1,DH2,DH3,DHJN,DdBHS,DHN2,Damour:2002et}. This simultaneous appearance of {\it billiards}
(with {\it chaotic} properties in important physical cases) and of an underlying
{\it symmetry} structure (infinite-dimensional Lie algebra) is an interesting
fact, which deserves to be studied in depth. This topic would have pleased Poincar\'e
because it involves several mathematical structures dear to his heart: notably, discrete
reflection groups (and their fundamental polytope), Lobachevskii space and chaotic dynamics.
Before explaining  the
techniques that have been used to uncover this connection, we will start by reviewing
 previous related works, and by stating the main results of this billiard/symmetry
 connection.

The simplest example of this connection concerns the
pure Einstein system in $D=3+1$-dimensional space-time. 
The Einstein's equations requiring the vanishing of the Ricci tensor ($R_{\m
\n}(g_{\a\b}) = 0)$ are non-linear PDE's for the metric components. Near a cosmological spacelike singularity, here chosen as $t=0$, the spatial gradients are expected
to become negligible compared to time derivatives (${\partial \over
\partial x^i } << {\partial \over
\partial t}$); this then suggests the decoupling of spatial points and allows
for an approximate treatment in which one replaces the above partial differential equations
by (a three-dimensional family) of ordinary differential equations.  Within this simplified context,
  Belinskii, Khalatnikov and Lifshitz (BKL)  gave a description \cite{BKL1,BKL2,BKL3} of
the asymptotic behavior of the general
solution of  the Einstein's equations, close to the singularity,  and showed that it can be described as a chaotic \cite{KLL,Bar}
sequence of generalized Kasner solutions. The Kasner metric is of the type
\beq g_{\a\b}(t) dx^{\a}dx^{\b} = - N^2 dt^2 + A_1 t^{2p_1} dx_1^2
+A_2 t^{2p_2} dx_2^2 +A_3 t^{2p_3} dx_3^2  \eeq where the
constants $p_i$ obey\footnote{In the $N=1$ gauge, they also obey $p_1+p_2+p_3=1$.} \beq \label{aux_metric}
\overrightarrow{p}^2 = p_1^2+p_2^2+p_3^2- (p_1+p_2+p_3)^2= 0. \eeq 
An exact Kasner solution, with a given set of $A_i$'s and $p_i$'s, can
be represented by a null line in a 3-dimensional auxiliary Lorentz space with
coordinates $p_1, p_2, p_3$ equipped with the metric given by the quadratic form
 $\overrightarrow{p}^2$ above. 
The auxiliary Lorentz space can be radially projected on the unit hyperboloid or further on the Poincar\'e disk
({\it i.e.} on the hyperbolic plane $H_2$):  the projection of a null line is a  geodesic on the hyperbolic plane.
See Figure 1.

\begin{figure}[ht]
\centerline{\includegraphics[scale=.8]{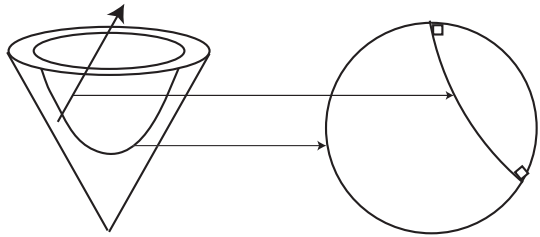}}
\caption{\small{Lorentz space and projection on Poincar\'e disk.}}
\end{figure}

BKL showed  that, because of non-linearities in  Einstein's equations,  the generic solution behaves as a succession of Kasner epochs, {i.e.},  as a broken null line in the auxiliary Lorentz space, or (after projection)
 a broken geodesic on the Poincar\'e disk. This broken geodesic motion is a ``billiard motion'' (seen either in Lorentzian
 space or in hyperbolic space). See Figure 2.

\begin{figure}[ht]
\centerline{\includegraphics[scale=.8]{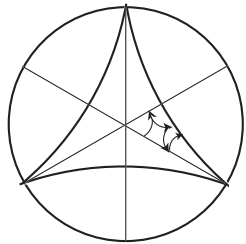}}
\caption{\small{Picture of chaotic cosmological behavior}}
\end{figure}

The billiard picture naturally follows from the Hamiltonian approach to cosmological behavior and was first obtained in the homogeneous (Bianchi IX)
four-dimensional case \cite {Chitre,Misnerb} and then extended to higher space-time dimensions with $p$-forms and dilatons \cite{Kirillov1993, KiMe,IvKiMe94,IvMe,AR,DH3,DHN2,DHRW,Damour:2002et}. Recent work
\cite{Damour:2002et} has improved the derivation of the billiard picture by using the
Iwasawa decomposition of the spatial metric.
Combining this decomposition with the Arnowitt-Deser-Misner \cite{ADM} Hamiltonian formalism highlights
 the mechanism by which all variables except the scale factors and the dilatons get asymptotically frozen.
 The non-frozen variables (logarithms of scale factors and dilatons) then undergo a billiard motion. This billiard
 motion can be seen either in  a $(D-1+n)$-dimensional Lorentzian space, or, after radial projection,
 on $(D-2+n)$-dimensional hyperbolic space. Here, $D$ is the spacetime dimension and $n$ the number
 of dilaton fields (see below for details). The Figures 1 and 2 correspond to the case $D=4$, $n=0$.
 \newline

A remarkable connection was also established \cite{DH1,DH2,DH3,DHJN,DdBHS,DHN2,Damour:2002et} between certain specific Einstein-matter systems and Lorentzian Kac-Moody (KM) algebras \cite{Kac}.  In the leading asymptotic approximation, this connection is simply that the Lorentzian billiard table within which the motion is confined can be identified with the  Weyl chamber of some corresponding Lorentzian KM algebra. This can happen only when many conditions are met: in particular, (i) the billiard table must be a Coxeter polyhedron (the dihedral angles between adacent walls must be integer submultiples of $\pi$) and ii) the billiard must be a simplex. Surprisingly, this occurs in many physically interesting Einstein-matter systems. For instance, pure Einstein gravity in $D$ dimensional space-time corresponds to the
Lorentzian KM algebra $AE_{D-1}$ \cite{DHJN} which is the overextension of the finite Lie algebra $A_{D-3}$:  for $D=4$, the algebra is $AE_3$ the Cartan
matrix of which is given by
\beq A= \left( \begin{array}{rrr} 2 & -1 & 0 \\
-1 & 2 &-2 \\  0 & -2& 2 \\  \end{array}  \right) \eeq Chaotic billiard tables have finite volume in hyperbolic space, while non-chaotic ones have infinite volume; as a consequence, chaotic billiards are associated with {\it hyperbolic} KM algebras; this happens to be the case for pure gravity when $D\leq 10$.

Another connection between physically interesting Einstein-matter systems and
KM algebras concerns the low-energy bosonic effective actions
arising in string and $M$ theories. Bosonic string theory in any space-time
dimension $D$ is related to the Lorentzian KM algebra $DE_D$
\cite{DH3,DdBHS}. The latter algebra is the canonical Lorentzian extension
of the finite-dimensional algebra $D_{D-2}$. The various superstring
theories (in the critical dimension $D=10$) and $M$-theory have been found
\cite{DH3} to be related either to $E_{10}$ (when there are two
supersymmetries in $D=10$, i.e. for type IIA, type IIB and $M$-theory) or
to $BE_{10}$ (when there is only one supersymmetry in $D=10$,
i.e. for type I and II heterotic theories), see the table below.
A construction of the Einstein-matter systems related  to the canonical Lorentzian extensions of
{\it all}  finite-dimensional Lie algebras $A_n$, $B_n$, $C_n$, $D_n$,
$G_2$, $F_4$, $E_6$, $E_7$ and $E_8$ (in
the above ``billiard'' sense) is presented in Ref.~\cite{DdBHS}.
See also Ref.~\cite{dBS} for the identification
of all hyperbolic KM algebras whose Weyl chambers are  Einstein billiards.

The correspondence between the
specific Einstein--three-form system (including a Chern--Simons term)
describing the bosonic sector of 11-dimensional supergravity (also known as
the ``low-energy limit of $M$-theory'') and the hyperbolic KM group
$E_{10}$ was studied in more detail in \cite{DHN2}. Reference  \cite{DHN2} introduces a formal expansion of the field
equations in terms of positive roots, i.e. combinations $\alpha =
\Sigma_i \, n^i \, \alpha_i$ of simple roots of $E_{10}$, $\alpha_i$, $i =
1,\ldots , 10$, where the $n^i$'s are integers $\geq 0$. It is then useful to
{\it order} this expansion according to the {\it height} of the positive root
$\alpha = \Sigma_i \, n^i \,
\alpha_i$, defined as ${\rm ht} (\alpha) = \Sigma_i \, n^i$. The correspondence
discussed above between the {\it leading} asymptotic evolution near a
cosmological singularity (described by a billiard) and Weyl chambers of
KM algebras involves only the terms in the field equation whose
height is ${\rm ht} (\alpha) \leq 1$. By contrast, the authors of Ref. \cite{DHN2} managed to show,
by explicit calculation, that there exists a way to define, at each
spatial point $x$, a correspondence between the field variables $g_{\mu\nu}
(t,x)$, $A_{\mu\nu\lambda} (t,x)$ (and their gradients), and a (finite)
subset of the parameters defining an element of the (infinite-dimensional)
coset space $E_{10} / K(E_{10})$ where $K(E_{10})$ denotes the maximal
compact subgroup of $E_{10}$, such that the (PDE) field equations of
supergravity get mapped onto the (ODE) equations describing a null geodesic
in $E_{10} / K(E_{10})$ {\it up to terms of height} 30. A complementary check of the
correspondence between $11$-dimensional supergravity and the $E_{10} / K(E_{10})$
$\sigma$-model has been obtained in \cite{DN04}. This result was further extended
to the correspondence between the $E_{10} / K(E_{10})$ $\sigma$-model and, both,
{\it massive} 10-dimensional IIA supergravity \cite{KN1},
and 10-dimensional IIB supergravity \cite{KN2}.

These tantalizing
results suggest that the infinite-dimensional hyperbolic Kac--Moody group $E_{10}$
may be a ``hidden symmetry'' of supergravity in the sense of mapping  
solutions onto solutions (the idea that $E_{10}$ might be
a symmetry of supergravity was first raised by
Julia long ago \cite{Julia,Julia2}). Note that the conjecture here is that the { \it continuous} group
$E_{10}(\RR)$ be a hidden symmetry group of { \it classical} supergravity.
At the {\it quantum} level, i.e. for M theory, one expects only a discrete version
of $E_{10}$, say $E_{10}(\ZZ)$, to be a quantum symmetry. See \cite{BGH}
for recent work on
extending the identification of \cite{DHN2} between roots of $E_{10}$ and symmetries
of supergravity/M-theory beyond height 30, and for references about previous suggestions
of a possible role for $E_{10}$. For earlier appearances of the Weyl groups
of the $E$ series in the context of $U$-duality see \cite{LPS,OPR,BFM}.
A series of recent papers  \cite{West,SWest,SWest2,Englert1,Englert2}
 has also explored the possible role of $E_{11}$
(a nonhyperbolic extension of $E_{10}$) as a hidden symmetry of M theory.

It is also tempting to assume that the
KM groups underlying the other (special) Einstein-matter systems
discussed above might be hidden (solution-generating) symmetries. For
instance, in the case of pure Einstein gravity in $D=4$ space-time, the
conjecture is that $AE_3$ be such a symmetry of Einstein gravity. This
case, and the correspondence between the field variables and the coset ones
is further discussed in \cite{Damour:2002et}.

Note that rigorous mathematical proofs \cite{AR, Rendall:2001nx, DHRW,IM02}
concerning the PDE/billiard connection are only available
for `non chaotic' billiards.\newline

In the remainder of this paper, we will outline various arguments explaining the above results;   a more complete derivation can be found in \cite{Damour:2002et}.

\subsection{General Models}
The general systems considered here are of the following form
\beq &&S[{\rm g}_{MN}, \phi, A^{(p)}] = \int d^D x \, \sqrt{- {\rm
g}} \;
\Bigg[R ({\rm g}) - \partial_M \phi \partial^M \phi \nonumber \\
&& \hspace{2.5cm} - \frac{1}{2} \sum_p \frac{1}{(p+1)!} e^{\l_p
\phi} F^{(p)}_{M_1 \cdots M_{p+1}} F^{(p)  \, M_1 \cdots M_{p+1}}
\Bigg] + \dots .~~~~~~~ \label{keyaction} \eeq Units are
chosen such that $16 \pi G_N = 1$,  $G_N$ is Newton's
constant and the space-time dimension $D \equiv d+1$ is left
unspecified. Besides the standard Einstein--Hilbert term the above
Lagrangian contains a dilaton\footnote{The generalization to any number of dilatons is
straightforward.}  field $\phi$ and a number of
$p$-form fields $A^{(p)}_{M_1 \cdots M_p}$ (for $p\geq 0$).  The $p$-form field strengths $F^{(p)} = dA^{(p)}$ are
normalized as \be F^{(p)}_{M_1 \cdots M_{p+1}} = (p+1)
\partial_{[M_1} A^{(p)}_{M_2 \cdots M_{p+1}]} \equiv
\partial_{M_1} A^{(p)}_{M_2 \cdots M_{p+1}} \pm p \hbox{
permutations }. \ee As a
convenient common formulation we adopt the Einstein conformal
frame and normalize the kinetic term of the dilaton $\phi$ with
weight one with respect to the Ricci scalar. The Einstein metric
${\rm g}_{MN}$ has Lorentz signature $(- + \cdots +)$ and is used
to lower or raise the indices; its determinant is denoted by ${\rm
g}$. The dots in the action (\ref{keyaction}) above
indicate possible modifications of the field strength by
additional Yang--Mills or Chapline--Manton-type couplings
\cite{pvnetal,CM}. The real parameter $\l_p$ measures the strength
of the coupling of $A^{(p)}$ to the dilaton. When $p=0$, we assume
that $\l_0\neq 0$ so that there is only one dilaton. \newline

\subsection{Dynamics in the vicinity of a spacelike singularity}

The main technical points that will be reviewed here are the following
\begin{itemize}
\item near the singularity, $t \rightarrow 0$, due to the decoupling of space points,  Einstein's PDE equations
 become ODE's
with respect to time. 

\item The study of these ODE's near $t\to 0$, shows that the $d \equiv D-1$ diagonal spatial metric components ``$g_{ii}$" and the dilaton $\phi$ move on a billiard in an auxiliary $d+1\equiv D$ dimensional Lorentz space.

\item All the
other field variables ($g_{ij}, i\neq j, A_{i_1...i_p}, \pi^{i_1...i_p})$ freeze as $t \rightarrow 0$. 

\item In many interesting cases, the billiard tables
can be identified with the fundamental Weyl chamber of an hyperbolic KM algebra.

\item For SUGRA$_{11}$, the KM algebra is $E_{10}$.  Moreover,
 the PDE's are equivalent  to
 the equations  of a null geodesic on the
coset space $E_{10} / K(E_{10}) $, up to height 30.
\end{itemize}

\subsubsection{Arnowitt-Deser-Misner Hamiltonian formalism}

To focus on the features relevant to the billiard picture, we
assume here that there are no Chern--Simons and no Chapline--Manton terms and that the
curvatures $F^{(p)}$ are abelian, $F^{(p)}
= d A^{(p)}$. That such additional terms do not alter the analysis has been proven in \cite{Damour:2002et}. In any pseudo-Gaussian gauge and in the temporal gauge ($g_{0i}=0$ and $A_{0 i_2...i_p}=0$, $\forall p$), the
Arnowitt-Deser-Misner Hamiltonian
action \cite{ADM} reads \beq && S\left[ g_{ij}, \pi^{ij}, \phi,
\pi_\phi, A^{(p)}_{j_1 \cdots j_p},
\pi_{(p)}^{j_1 \cdots j_p}\right] = \nonumber \\
&& \hspace{1cm} \int dx^0 \int d^d x \left( \pi^{ij} \dot{g_{ij}}
+ \pi_\phi \dot{\phi} + \frac{1}{p!}\sum_p \pi_{(p)}^{j_1 \cdots
j_p} \dot{A}^{(p)}_{j_1 \cdots j_p} - H \right)\,,
\label{GaussAction} \eeq  where the Hamiltonian density $H$ is
\beq\label{Ham}
H &\equiv&  \tilde{N} \ch \, ,\\[2mm]
\label{Ham1}\ch &=& \ck + \cm \, ,\\[2mm]
\ck &=& \pi^{ij}\pi_{ij} - \frac{1}{d-1} \pi^i_{\;i} \pi^j_{\;j}
+ \frac{1}{4} \pi_\phi^2 
+ \sum_p \frac{e^{- \lambda_p \phi}} {2 \, p!} \, \pi_{(p)}^{j_1
\cdots j_p}
\pi_{(p) \, j_1 \cdots j_p} \, ,~~~~~\\[2mm]
\cm &=& - g R + g g^{ij} \partial_i \phi \partial_j \phi + \sum_p
\frac{e^{ \lambda_p \phi}}{2 \; (p+1)!} \, g \, F^{(p)}_{j_1
\cdots j_{p+1}} F^{(p) \, j_1 \cdots j_{p+1}}\,, \eeq

\noindent and $R$ is the spatial curvature scalar. $\tilde{N} = N/\sqrt{g^{(d)}}$ is the rescaled lapse. The dynamical
equations of motion are obtained by varying the above action with
respect to the spatial metric components, the dilaton, the spatial
$p$-form components and their conjugate momenta. In addition,
there are constraints on the dynamical variables,

\beq
\ch &\approx& 0  \; \; \; \; \; \; \hbox{(``Hamiltonian constraint")}, \\[2mm]
\ch_i &\approx& 0  \; \; \; \; \; \; \hbox{(``momentum constraint")}, \\[2mm]
\varphi_{(p)}^{j_1 \cdots j_{p-1}} &\approx& 0 \; \; \; \;\; \;
\hbox{(``Gauss law" for each $p$-form), } \label{Gauss} \eeq with
\beq \ch_i &=& -2 {\pi^j}_{i|j} + \pi_\phi
\partial_i \phi + \sum_p \frac1{p!} \
\pi_{(p)}^{j_1 \cdots j_p} F^{(p)}_{i j_1 \cdots j_{p}} \,,\\[2mm]
\varphi_{(p)}^{j_1 \cdots j_{p-1}} &=& {\pi_{(p)}^{j_1 \cdots
j_{p-1} j_p}}_{\vert j_p}\,, \eeq where the subscript $|j$ stands
for spatially covariant derivative. \newline

\subsubsection{Iwasawa decomposition of the spatial metric}

 We systematically use the Iwasawa decomposition of the spatial
metric $g_{ij}$ and write \be \label{Iwasawaex} g_{ij} =
\sum_{a=1}^d e^{- 2 \b^a} {\cn^a}_i  \, {\cn^a}_j \ee where $\cn$
is an upper triangular matrix with $1$'s on the diagonal.
We will also need the Iwasawa coframe $\{ \theta^a \}$,
\be\label{Iwasawa1} \theta^a = {\cn^a}_i \, dx^i\,, \ee as well as
the vectorial frame $\{ e_a \}$ dual to the coframe $\{ \theta^a
\}$, \be\label{Iwasawa2} e_a = {\cn^i}_a \frac{\partial}{\partial
x^i} \ee where the matrix ${\cn^i}_a$ is the inverse of
${\cn^a}_i$, i.e., ${\cn^a}_i {\cn^i}_b = \delta^a_b$.  It is
again an upper triangular matrix with 1's on the diagonal. Let us
now examine how the Hamiltonian action gets transformed when one
performs, at each spatial point, the Iwasawa decomposition
\Ref{Iwasawaex} of the spatial metric. The kinetic terms of the
metric and of the dilaton in the Lagrangian (\ref{keyaction}) are
given by the quadratic form \be\label{dsigma2} G_{\mu\nu}d\beta^\mu d\beta^\nu =\sum_{a=1}^d (d
\beta^a)^2 - \left(\sum_{a=1}^d d \beta^a\right)^2  + d\phi^2 , Ê\quad \beta^\mu = (\beta^a,\phi).\ee The change of variables $(g_{ij}\to \beta^a, {\cn^a}_i )$ corresponds
to a point transformation and can be extended to the momenta as a canonical transformation in the standard way via \be\label{cantra}
\p^{ij}\dot{g}_{ij} \equiv \sum_a \pi_a \dot{\b}^a + \sum_{a}
{P^i}_a \dot{{\mathcal N}^a}_{i} \,\,. \ee Note that the momenta
\be\label{Nmomenta} {P^i}_a = \frac{\partial\mathcal L}{\partial
\dot{{\mathcal N}^a}_i} = \sum_{b} e^{2(\beta^b - \beta^a)}
{\dot\cn^a}_{\;\;j} {\cn^j}_b {\cn^i}_b \ee conjugate to the
nonconstant off-diagonal Iwasawa components ${\cn^a}_i$ are only
defined for $a<i$; hence the second sum in (\ref{cantra}) receives
only contributions from $a<i$.\newline

\subsubsection{Splitting of the Hamiltonian}

We next split the Hamiltonian  density\, $\ch$  (\ref{Ham}) in two
parts: ${\mathcal H}_0$, which is the kinetic term
for the local scale factors and the dilaton $\beta^\mu= (\beta^a, \phi)$, and
$\cv$, a ``potential density'' (of
weight 2) , which contains everything else. Our
analysis below will show why it makes sense to group the kinetic
terms of both the off-diagonal metric components and the $p$-forms
with the usual potential terms, i.e. the term $\mathcal M$ in
(\ref{Ham1}).  Thus, we write \be \ch =  {\mathcal H}_0 + \cv
\label{HplusV} \ee with the kinetic term of the $\b$ variables
\be\label{eq3.23} {\mathcal H}_0 = \frac{1}{4}\, G^{\mu\nu}
\pi_\mu \pi_\nu\,, \ee where $G^{\mu\nu}$ denotes the inverse of
the metric $G_{\mu\nu}$ of Eq.~(\ref{dsigma2}). In other words,
the right hand side of Eq.~(\ref{eq3.23}) is defined by
\begin{equation}\label{Gmunuup}
G^{\mu \nu} \pi_\mu \pi_\nu \equiv \sum_{a=1}^d \pi_a^2 -
\frac{1}{d-1} \left(\sum_{a=1}^d \pi_a\right)^2 + \pi_\phi^2\,,
\end{equation}
where $ \pi_\mu \equiv (\pi_a, \pi_\phi)$ are the momenta
conjugate to $\beta^a$ and $\phi$, respectively, i.e. \be \pi_\m =
2 \tilde{N}^{-1} G_{\m \n} \dot{\beta}^\n = 2 G_{\m \n} \frac { d
{\beta}^\n}{d\tau}\, . \ee The total (weight 2) potential density,
\be \cv = \cv_S + \cv_G + \sum_p \cv_{p}  + \cv_\phi\, , \ee is
naturally split into a ``centrifugal'' part $\cv_S$ linked to the kinetic
energy of the off-diagonal components (the index $S$ referring to
``symmetry,''), a ``gravitational'' part $\cv_G$, a term
from the $p$-forms, $\sum_p \cv_{p}$, which is a sum of an ``electric'' and a
``magnetic'' contribution and also a  contribution to
the potential coming from the spatial gradients of the dilaton
$\cv_\phi$.

\begin{itemize}

\item{``centrifugal'' potential}
\be \label{centrifugal} \cv_S = \frac{1}{2} \sum_{a<b}
e^{-2(\beta^b - \beta^a)} \left( {P^j}_b {{\mathcal
N}^a}_j\right)^2, \ee

\item{ ``gravitational'' (or ``curvature'') potential}

\be \label{gravitational} \cv_G =  - g R\, = \frac{1}{4}
{\sum_{a\neq b \neq c}} e^{-2\a_{abc}(\beta)} (C^a_{\; \; bc})^2 -
\sum_a e^{-2 \m_a(\beta)} F_a\,, \ee where  \be \a_{abc}(\beta)\equiv \sum_e \beta^e + \beta^a-\beta^b-\beta^c,\, a\neq b, b\neq c, c\neq a\ee and \be d\theta^a =-\frac{1}{2}C^a_{\; \; bc}\theta^b\wedge\theta^c\ee while $F_a$ is a polynomial of
degree two in the first derivatives $\partial \beta$ and of degree one
in the second derivatives $\partial^2 \beta$. 

\item{$p$-form potential}
\be
 \cv_{(p)} = \cv_{(p)}^{el} + \cv_{(p)}^{magn}\,,
\ee which is a sum of an ``electric'' $\cv_{(p)}^{el}$ and a
``magnetic'' $\cv_{(p)}^{magn}$ contribution. The ``electric''
contribution can be written as \beq \cv_{(p)}^{el} &=& \frac{e^{-
\lambda_p \phi}} {2 \,
p!} \, \pi_{(p)}^{j_1 \cdots j_p} \pi_{(p) \, j_1 \cdots j_p} \nonumber\\
&=& \frac{1}{2 \, p!} \sum_{a_1, a_2, \cdots, a_p} e^{-2 e_{a_1
\cdots a_p}(\b)} (\ce^{a_1  \cdots a_p})^2 \,,\eeq where $
\ce^{a_1 \cdots a_p} \equiv {\cn^{a_1}}_{j_1} {\cn^{a_2}}_{j_2}
\cdots {\cn^{a_p}}_{j_p} \pi^{j_1 \cdots j_p}\,,$  and $e_{a_1
\cdots a_p}(\b)$ are the "electric wall" forms, \be e_{a_1 \cdots
a_p}(\b) = \b^{a_1} + \cdots + \b^{a_p} + \frac{\l_p}{2} \phi  \,.
\ee And the ``magnetic'' contribution reads,

\beq \cv_{(p)}^{magn} &=&  \frac{e^{ \lambda_p \phi}}{2 \; (p+1)!}
\, g \, F^{(p)}_{j_1 \cdots j_{p+1}} F^{(p) \, j_1 \cdots j_{p+1}}
\nonumber\\&=&
\frac{1}{2 \, (p+1)!} \sum_{a_1, a_2, \cdots, a_{p+1}} e^{-2
m_{a_{1}  \cdots a_{p+1}}(\b)} (\cf_{a_1  \cdots a_{p+1}})^2 \,.
\eeq where $ \cf_{a_1 \cdots a_{p+1}} = {\cn^{j_1}}_{a_1} \cdots
{\cn^{j_{p+1}}}_{a_{p+1}} F_{j_1 \cdots j_{p+1}}\, $ and the
$m_{a_{1} \cdots a_{p+1}}(\b)$ are the magnetic linear forms \be
m_{a_{1} \cdots a_{p+1}}(\b) = \sum_{b \notin \{a_1,a_2,\cdots
a_{p+1}\}} \!\b^b - \frac{\l_p}{2}\, \phi\,, \ee
\item{dilaton potential}
\beq \cv_\phi  &=& g g^{ij}
\partial_i \phi \partial_j \phi\ \\
&=& \sum_a e^{-\mu_a(\beta)} (\mathcal{N}_a{}^i \partial_i
\phi)^2,. \eeq where \beq \mu_a(\b) = \sum_e \b^e - \b^a \eeq

\end{itemize}

\subsubsection{Appearance of sharp walls in the BKL limit}

In the decomposition of the hamiltonian as $\mathcal{H} = \mathcal{H}_0 +
\cv$, $\mathcal{H}_0$ is the kinetic term for the $\beta^\mu$'s while all other variables now only appear through the potential $\cv$ which is
schematically of the form \beq \label{V1} \cv(
 \b^\m, \partial_x \b^{\m}, P,Q)
=\sum_A c_A( \partial_x \b^{\m}, P,Q) \exp\big(-
2 w_A (\beta) \big)\,, \eeq  where $(P,Q) = ({\cn^a}_i, {P^i}_a,
\ce^{a_1 \cdots a_p},\cf_{a_1 \cdots a_{p+1}})$. Here $w_A (\beta)
= w_{A \m} \b^\m$ are the linear wall forms already introduced above: \beq \mbox{symmetry walls}&:& w^S_{ab}\equiv \beta^b - \beta^a; \quad a<b\nonumber\\ \mbox{gravitational walls}&:& \a_{abc}(\b) \equiv \sum_e \beta^e + \beta^a-\beta^b-\beta^c,\, a\neq b, b\neq c, c\neq a\nonumber\\ &\,& \m_a(\beta)\equiv\sum_e \beta^e -\beta^a,\nonumber\\ \mbox{electric walls}&:& e_{a_1 \cdots a_p}(\b)\equiv\beta^{a_1}+...+\beta^{a_p} + \frac{1}{2}\lambda_p\phi,\nonumber\\ \mbox{magnetic walls}&:& m_{a_1 \cdots a_{p+1}}(\b)\equiv\sum_e \beta^e -\beta^{a_1}-...-\beta^{a_{p+1}}-   \frac{1}{2}\lambda_p\phi. \nonumber \eeq  In order to take the limit $t\rightarrow 0$ which corresponds to
$\b^{\m} $ tending to future time-like infinity, we decompose $\b^{\m}$
into hyperbolic polar coordinates $(\rho,\g^{\m})$, i.e. \beq
\b^{\m} = \rho \g^{\m} \eeq where $\g^{\m}$ are coordinates on the
future sheet of the unit hyperboloid which are constrained by \beq
G_{\m\n} \g^{\m} \g^{\n} \equiv \g^{\m} \g_{\m} = -1\eeq and
$\rho$ is the time-like variable defined by \beq \rho^2 \equiv - G_{\m \n}
\b^{\m} \b^{\n} \equiv - \b_{\m} \b^{\m} >0, \eeq which behaves like $\rho\sim -\ln t \to +\infty$ at the BKL limit. In terms of these
variables, the potential term looks like

\beq \sum_A c_A(  \partial_x \b^{\m}, P,Q) \rho^2 \exp\big(- 2
\rho w_A (\g) \big)\,. \eeq
The essential point now is that, since $\r \to +
\infty$, each term $\rho^2 \exp\big(- 2 \rho w_A (\gamma) \big)$
becomes a {\it sharp wall potential}, i.e. a function of  $w_A
(\gamma)$ which is zero when $w_A (\gamma) >0$, and  $+\infty$
when $w_A (\gamma) < 0$. To formalize this behavior we define the
sharp wall $\Theta$-function\,\footnote{\,One should more properly
write $\Theta_\infty(x)$, but since this is the only step function
encountered here, we use the simpler notation
$\Theta(x)$.} as \be \Theta (x) := \left\{ \begin{array}{ll}
                      0  & \mbox{if $x<0$} \,,\\[1mm]
                      +\infty & \mbox{if $x>0$}\,.
                      \end{array}
                      \right.
\ee A basic formal property of this $\Theta$-function is its
invariance under multiplication by a positive quantity. Because
all the relevant prefactors $c_A( \partial_x \b^{\m}, P,Q)$ are
generically {\it positive} near each leading wall, we can formally
write
\begin{eqnarray}
\lim_{\rho\rightarrow\infty} && \Big[ c_A( \partial_x \b^{\m},Q,P) \rho^2
\exp\big(-\rho w_A (\gamma) \Big] = c_A(Q,P)\Theta\big(-2
w_A (\gamma) \big) \nonumber\\[2mm]
&\equiv &\Theta\big(- 2 w_A (\gamma) \big) \, 
\end{eqnarray}
valid in spite of the increasing of the spatial gradients \cite{Damour:2002et}.
Therefore, the limiting dynamics is equivalent to a free motion in
the $\b$-space interrupted by reflections against hyperplanes in
this $\b$-space given by $w_A (\b) = 0$ which correspond to a
potential described by infinitely high step functions

\beq \cv(\b, P,Q) = \sum_A \Theta\big(-2 w_A (\g) \big) \eeq The
other dynamical variables (all variables but the $\b^\mu$'s) completely disappear from this
limiting Hamiltonian and therefore they all get frozen as $t\rightarrow
0$.
\subsubsection{Cosmological singularities and Kac--Moody algebras}

Two kinds of motion are possible according to the volume of the
billiard table on which it takes place, i.e. the volume of the region where $\cv = 0$ for $t\to 0$, also characterized by the conditions,

\beq w_A(\b) > 0 \quad \forall A . \eeq
Depending on the fields present in the lagrangian, on their dilaton-couplings and on the spacetime dimension, the
(projected) billiard volume is either of finite or infinite. The
finite volume case corresponds to never-ending, chaotic oscillations for the $\beta$'s
while in the infinite volume
case, after a finite number of reflections off the walls, they tend to
an asymptotically monotonic Kasner-like behavior, see Figure 3:

\begin{figure}[ht]
\centerline{\includegraphics[scale=.8]{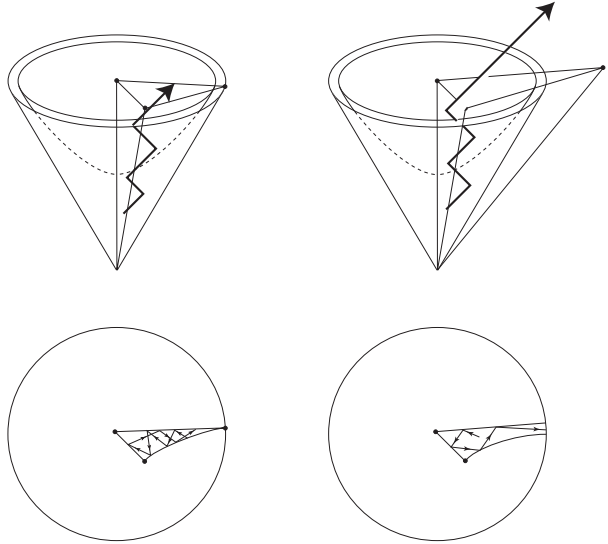}}
\caption{\small{Sketch of billiard tables describing the asymptotic cosmological behavior of Einstein-matter systems.}}
\end{figure} {\small{In Figure 3 the upper panels
are drawn in the Lorentzian space spanned by $(\beta^{\mu}) =
(\beta^a , \phi)$. The billiard tables are represented as
``wedges'' in $(d+1)$--dimensional (or $d$-dimensional, if there are
no dilatons) $\beta$-space, bounded by
hyperplanar walls $w_A (\beta) = 0$ on which the
billiard ball undergoes spe\-cular reflections. The upper left
panel is a  (critical) ``chaotic'' billiard table (contained
within the $\beta$-space future  light cone), while the upper
right one is a (subcritical) ``nonchaotic'' one (extending beyond
the light cone). The lower panels represent the corresponding
billiard tables (and billiard motions) after projection onto
hyperbolic space  $H_d$ ($H_{d-1}$ if there are no dilatons). The latter
projection is defined in the text by central projection onto
$\gamma$-space ({i.e.} the unit hyperboloid $G_{\mu\nu} \,
\gamma^{\mu} \, \gamma^{\nu} = -1$, see the upper panels), and is
represented in the lower panels by its image in the Poincar\'e
ball (disk). }}\newline


In fact, not all the walls are relevant for determining
the billiard table. Some of the walls stay behind the
others and are not met by the billiard ball. Only a subset of the walls $w_A(\b)$, called
dominant walls and here denoted $\{w_i(\b)\}$ are needed to delimit the hyperbolic domain. Once
the dominant walls $\{w_i(\b)\}$ are found, one can compute the following matrix
\beq A_{ij} \equiv 2 { w_i.w_j \over w_{i}.  w_{i} } \eeq where
$w_i . w_j = G^{\m\n} w_{i\m} w_{j\n }$. By definition,  the diagonal elements  are all equal to 2. Moreover, in many interesting cases, the off-diagonal elements happen to be non positive integers. These are precisely the characteristics of a generalized Cartan matrix, namely that  of an infinite KM algebra (see appendix A). As
recalled in the introduction, for pure gravity in $D$ space-time dimensions, there are $D-1$ dominant
walls and the
matrix $A_{ij}$ is exactly the generalized Cartan matrix of the hyperbolic KM
algebra $AE_{D-1} \equiv A^{\wedge\wedge}_{D-3} \equiv A^{++}_{D-3}$  which is hyperbolic for $D\leq10$. More generally, bosonic string theory in $D$
space-time dimensions is related to the Lorentzian
KM algebra $DE_D$ \cite{DH3,DdBHS} which is the canonical
Lorentzian extension of the finite-dimensional Lie algebra $D_{D-2}$.
The various superstring theories, in the critical dimension
$D=10$, and $M$-theory have been found \cite{DH3} to be related
either to $E_{10}$ (when there are two supersymmetries,
i.e. for type IIA, type IIB and $M$-theory) or to $BE_{10}$ (when
there is only one supersymmetry, i.e. for type I and II
heterotic theories), see the table.\newline

\noindent The hyperbolic KM algebras are those relevant for chaotic billiards since their fundamental Weyl chamber has a finite volume. \newline

\begin{centering}
\begin{tabular}{|c|p{6.5cm}|}
\hline Theory & Corresponding Hyperbolic KM algebra \\  \hline Pure gravity in
$D \leq 10$ & \scalebox{.5} {
\begin{picture}(180,60)
\put(5,-5){$\alpha_{1}$} \put(45,-5){$\alpha_2$}
 \put(125,-5){$\alpha_3$}  \put(50,45){$\alpha_{D-1}$}\put(85,-5){$\alpha_4$}
  \put(140,45){$\alpha_5$}
\thicklines \multiput(10,10)(40,0){4}{\circle{10}}
\multiput(15,10)(40,0){3}{\line(1,0){30}}
\multiput(90,50)(40,0){2}{\circle{10}}
\put(130,15){\line(0,1){30}} \put(50,15){\line(1,1){35}}
\dashline[0]{2}(95,50)(105,50)(115,50)(125,50)
\end{picture}
}
 \\ \hline
 M-theory, IIA and  IIB Strings & \scalebox{.5}{
\begin{picture}(180,60)
\put(5,-5){$\alpha_{1}$} \put(45,-5){$\alpha_2$}
\put(85,-5){$\alpha_3$}
 \put(125,-5){$\alpha_4$}
  \put(165,-5){$\alpha_5$} \put(205,-5){$\alpha_6$}
  \put(245,-5){$\alpha_7$}   \put(285,-5){$\alpha_8$}
  \put(325,-5){$\alpha_9$}
  \put(260,45){$\alpha_{10}$}
\thicklines \multiput(10,10)(40,0){9}{\circle{10}}
\multiput(15,10)(40,0){8}{\line(1,0){30}}
\put(250,50){\circle{10}} \put(250,15){\line(0,1){30}}
\end{picture}
 }
 \\ \hline
 type I and heterotic Strings & \scalebox{.5}{
\begin{picture}(180,60)
\put(5,-5){$\alpha_{1}$}
\put(45,-5){$\alpha_2$}\put(85,-5){$\alpha_3$}
 \put(125,-5){$\alpha_{4}$}
  \put(165,-5){$\alpha_{5}$}
\put(205,-5){$\alpha_{6}$} \put(245,-5){$\alpha_{7}$}
\put(285,-5){$\alpha_{8}$} \put(325,-5){$\alpha_{9}$}
  \put(70,45){$\alpha_{10}$}
\thicklines \multiput(10,10)(40,0){9}{\circle{10}}
\multiput(15,10)(40,0){7}{\line(1,0){30}}
\dashline[0]{2}(95,10)(105,10)(115,10)(125,10)
\put(295,7.5){\line(1,0){30}}\put(295,12.5){\line(1,0){30}}
\put(305,0){\line(1,1){10}} \put(305,20){\line(1,-1){10}}
\put(90,50){\circle{10}} \put(90,15){\line(0,1){30}}
\end{picture}
} \\ \hline
closed bosonic string in $D=10$ & \scalebox{.5}{
\begin{picture}(180,60)
\put(5,-5){$\alpha_{1}$} \put(45,-5){$\alpha_2$}
\put(85,-5){$\alpha_3$}
 \put(125,-5){$\alpha_4$}
  \put(165,-5){$\alpha_5$} \put(205,-5){$\alpha_6$}
  \put(245,-5){$\alpha_7$}   \put(285,-5){$\alpha_8$}
  \put(100,45){$\alpha_9$}
  \put(260,45){$\alpha_{10}$}
\thicklines \multiput(10,10)(40,0){8}{\circle{10}}
\multiput(15,10)(40,0){7}{\line(1,0){30}}
\put(250,50){\circle{10}} \put(250,15){\line(0,1){30}}
\put(90,50){\circle{10}} \put(90,15){\line(0,1){30}}
\end{picture} }\\ \hline
\end{tabular}
\end{centering}\newline

\noindent {\small{This table displays the Coxeter--Dynkin diagrams which encode the geometry of the
billiard tables describing the asymptotic cosmological behavior of General Relativity and of
three blocks of string theories: ${\mathcal B}_2 = \{$$M$-theory,
type IIA and type IIB superstring theories$\}$, ${\mathcal B}_1 =
\{$type I and the two heterotic superstring theories$\}$, and
${\mathcal B}_0 = \{$closed bosonic string theory in $D=10\}$.
Each node of the diagrams represents a dominant wall of the
cosmological billiard. Each Coxeter diagram of a
billiard table corresponds to the Dynkin diagram of a (hyperbolic) KM algebra: $E_{10}$,
$BE_{10}$ and $DE_{10}$}} .\newline

\noindent The precise links between a chaotic billiard and its corresponding Kac--Moody
algebra can be summarized as follows
\begin{itemize}
\item the scale factors $\b^{\mu}$ parametrize a Cartan element $h = \sum_{\m=1}^{r} \b^{\m}h_{\mu}
$,
\item  the dominant walls $w_i(\b), (i=1,...,r)$ correspond to the simple roots
$\a_i$ of the KM algebra,
 \item the group of reflections in the
cosmological billiard is the Weyl group of the KM algebra,
and
\item the billiard table can be identified with the Weyl chamber
of the KM algebra.
\end{itemize}

\subsubsection{$E_{10} $ and a ``small tension" limit of SUGRA$_{11}$}

The main feature of the gravitational billiards that can be
associated with the KM algebras is that there exists a group
theoretical interpretation of the billiard motion: the asymptotic
BKL dynamics is equivalent (in a sense to be made precise below),
at each spatial point, to the asymptotic dynamics of a
one-dimensional nonlinear $\sigma$-model based on a certain
infinite-dimensional coset space $G/K$, where the KM group $G$ and
its maximal compact subgroup $K$ depend on the specific model. As
we have seen, the walls that determine the billiards are the {\it
dominant walls}. For the KM billiards, they correspond to the {\it
simple roots} of the KM algebra. As we discuss below, some of the
subdominant walls also have an algebraic interpretation in terms
of higher-height positive roots. This enables one to go beyond the
BKL limit and to see the beginnings of a possible identification
of the dynamics of the scale factors {\em and} of all the
remaining variables with that of a nonlinear $\sigma$-model
defined on the  cosets of the KM group divided by its
maximal compact subgroup \cite{DHN2,Damour:2002et}.\newline


For concreteness, we will only consider one specific example here:
the relation between the cosmological evolution of $D=11$
supergravity and a null geodesic on $E_{10} / K(E_{10})$
\cite{DHN2} where $KE_{10}$ is the maximally compact subgroup of  $E_{10}$. The $\sigma$-model is formulated in terms of a one-parameter dependent group element $\cv=\cv(t)\in E_{10}$ and its Lie algebra value derivative \be v(t) :=\frac{d\cv}{dt}\cv^{-1}(t) \in e_{10}.\ee The action is \be S_1^{E_{10}} = \int{dt \over n(t)} <v_{sym}(t)\vert v_{sym}(t)> \ee with a lapse function $n(t)$ whose variation gives rise to the Hamiltonian constraint ensuring that the trajectory is a null geodesic. The symmetric projection \be v_{sym} := \frac{1}{2} (v+v^T)\ee  is introduced in order to define an evolution on the coset space. 
Here $< . \vert . >$
is the standard invariant bilinear form on $E_{10}$ ; $v^T$ is the ``transpose"
of $v$ defined with the Chevalley involution\footnote{The Chevalley involution is defined by $\omega(h_i ) =-h_i ; \
\omega(e_i ) =-f_i ; \ \omega(f_i ) = -e_i $} 
as $v^T =- \omega(v)$. This action is
invariant under $E_{10}$:

\beq \cv(t) \rightarrow k(t) \cv(t) g \hspace{1cm} \mbox{where}\hspace{1cm} k \in KE_{10} \
g\in E_{10} \eeq Making use of the explicit Iwasawa parametrization of
the generic $E_{10}$ group element $\cv = K A N$ together with the gauge choice $K= 1$ (Borel gauge), one can write
$$ {{\mathcal V}}(t) = \exp X_h (t) \cdot \exp X_A (t)$$
with $X_h(t) = {h^a}_b  {K^b}_a$ and $$X_A (t) = \ft1{3!} A_{abc}
E^{abc} + \ft1{6!} A_{a_1\dots a_6} E^{a_1\dots a_6} + \ft1{9!}
A_{a_0|a_1 \dots a_8} E^{a_0|a_1 \dots a_8} + \dots\, .$$ Using
the $\E$ commutation relations in $GL(10)$ form
(see \cite{KN1,KN2} for other decompositions of the $\E$ algebra)
together with the
bilinear form for $\E$, one obtains up to height 30 \footnote{We keep
only the generators $E^{abc}, \ E^{a_1\dots a_6}$ and $E^{a_0|a_1
\dots a_8}$ corresponding to the $E_{10}$ roots $\a = \sum n_i \a_i $ with height $\sum_i
n_i \leq 29 $ ($\a_i$ are simple roots and $n_i$ integers)} ,

\begin{eqnarray}\label{Lag}
n {{\mathcal L}} &=& \ft14 (g^{ac} g^{bd} - g^{ab} g^{cd}) \dot
g_{ab} \dot g_{cd}
  + \ft12 \ft1{3!} DA_{a_1a_2a_3}DA^{a_1a_2a_3} \non[2mm]
&&  
 + \ft12 \ft1{6!} DA_{a_1 \dots a_6}DA^{a_1\dots a_6}
  + \ft12 \ft1{9!} DA_{a_0 | a_1 \dots a_8} DA^{a_0|a_1\dots a_8}\,,
\end{eqnarray}
where $g^{ab} = {e^a}_c {e^b}_c$ with $$ {e^a}_b \equiv {(\exp
h)^a}_b\,,$$ and all ``contravariant indices'' have been raised by
$g^{ab}$. The ``covariant'' time derivatives are defined by (with
$\partial A\equiv \dot A$)
\begin{eqnarray}\label{Dtime}
DA_{a_1a_2a_3} &:=& \partial A_{a_1a_2 a_3} \,,\nonumber\\[2mm]
DA_{a_1\dots a_6} &:=& \partial A_{a_1 \dots a_6}
    + 10 A_{[a_1a_2 a_3} \partial A_{a_4a_5 a_6]} \,,\non[2mm]
DA_{a_1|a_2\dots a_9} &:=& \partial A_{a_1|a_2 \dots a_9}
    + 42 A_{\langle a_1a_2 a_3} \partial A_{a_4 \dots  a_9 \rangle} \non[2mm]
&& 
- 42 \partial A_{\langle a_1a_2 a_3} A_{a_4 \dots  a_9 \rangle}
    + 280 A_{\langle a_1a_2 a_3} A_{a_4a_5 a_6} \partial A_{a_7a_8
a_9\rangle} \, .~~~~~~~~~
\end{eqnarray}
Here antisymmetrization $[\dots]$, and projection on the $\ell =
3$ representation $\langle \dots \rangle$, are  normalized with
strength one (e.g. $[[\dots]] = [\dots]$). Modulo field
redefinitions, all numerical coefficients in \Ref{Lag} and in
\Ref{Dtime} are uniquely fixed by the structure of $\E$.\newline

In order to compare the above coset model results with those of the bosonic part of $D=11$ supergravity, we recall the action
\begin{eqnarray}
\label{sugra} S^{sugra}_{11} &= &\int d^{11} x \Bigl[ \sqrt{-{\rm G}} \, R({\rm
G}) - \frac{\sqrt{-{\rm G}}}{48} \, {\mathcal
F}_{\alpha\beta\gamma\delta} \, {\mathcal
F}^{\alpha\beta\gamma\delta} \nonumber \\
&&+ \frac{1}{(12)^4} \, \varepsilon^{\alpha_1 \ldots \alpha_{11}}
\, {\mathcal F}_{\alpha_1 \ldots \alpha_4} \, {\mathcal
F}_{\alpha_5 \ldots \alpha_8} \, {\mathcal A}_{\alpha_9
\alpha_{10} \alpha_{11}} \Bigl]\,.
\end{eqnarray}
The space-time indices $\alpha , \beta , \ldots $  take the values  $ 0,1,
\ldots , 10$;  $\varepsilon^{01 \ldots 10} = +1$, and 
the four-form ${\mathcal F}$ is the exterior derivative of
${\mathcal A}$, ${\mathcal F} = d {\mathcal A}$. Note the presence
of the Chern--Simons term ${\mathcal F} \wedge {\mathcal F} \wedge
{\mathcal A}$ in the action (\ref{sugra}). Introducing a
zero-shift slicing ($N^i=0$) of the eleven-dimensional space-time,
and a {\em time-independent} spatial zehnbein $\theta^a(x) \equiv
{E^a}_i(x) dx^i$, the metric and four-form ${{\mathcal F}} = d\cA$
become
\begin{eqnarray}\label{Gauge}
  &&  ds^2 = {\rm G}_{\alpha\beta} \, dx^{\alpha} \, dx^{\beta} = - N^2
(d{x^0})^2 + G_{ab} \theta^a \theta^b \\ \!\!\!\!\!\! {{\mathcal
F}} &=& \frac1{3!}{{\mathcal F}}_{0abc}\,  dx^0
\!\wedge\!\theta^a\!\wedge\! \theta^b\!\wedge\! \theta^c +
\frac1{4!}{{\mathcal F}}_{abcd} \,
\theta^a\!\wedge\!\theta^b\!\wedge\! \theta^c\!\wedge\!\theta^d .
\nn
\end{eqnarray}
We choose the time coordinate $x^0$ so that the lapse
$N=\sqrt{G}$, with $G:= \det G_{ab}$ (note that $x^0$ is not the
proper time\,\footnote{\,In this section, the proper time is denoted
by $T$ while the variable $t$ denotes the parameter of the
one-dimensional $\sigma$-model introduced above.} $T = \int N
dx^0$; rather, $x^0\rightarrow\infty$ as $T \rightarrow 0$). In
this frame the complete evolution equations of $D=11$ supergravity
read
\begin{eqnarray}\label{EOM}
\partial_0 \big( G^{ac} \partial_0 G_{cb} \big)  &=&
\ft16 G {{\mathcal F}}^{a\beta\gamma\delta} {{\mathcal
F}}_{b\beta\gamma\delta} - \ft1{72} G {{\mathcal
F}}^{\alpha\beta\gamma\delta} {{\mathcal
F}}_{\alpha\beta\gamma\delta} \delta^a_b 
- 2 G {R^a}_b (\Gamma,C)\,, \nonumber\\[2mm]
\partial_0 \big( G{{\mathcal F}}^{0abc}\big) &=&
\ft1{144} \varepsilon^{abc a_1 a_2 a_3 b_1 b_2 b_3 b_4}
          {{\mathcal F}}_{0a_1 a_2 a_3} {{\mathcal F}}_{b_1 b_2 b_3 b_4} \nonumber\\[1mm]
  &&  
 + \ft32 G {{\mathcal F}}^{de[ab} {C^{c]}}_{de} - G {C^e}_{de} {{\mathcal F}}^{dabc}
     - \partial_d \big( G{{\mathcal F}}^{dabc} \big)\,, \nonumber\\[2mm]
\partial_0 {{\mathcal F}}_{abcd} &=& 6 {{\mathcal F}}_{0e[ab} {C^e}_{cd]} + 4
\partial_{[a} {{\mathcal F}}_{0bcd]}\,,
\end{eqnarray}
where $a,b \in \{1,\dots,10\}$ and $\alpha,\beta \in
\{0,1,\dots,10\}$, and $R_{ab}(\Gamma,C)$ denotes the spatial
Ricci tensor; the (frame) connection components are given by $ 2
G_{ad} {\Gamma^d}_{bc} = C_{abc} + C_{bca} - C_{cab} +
          \partial_b G_{ca} + \partial_c G_{ab} - \partial_a G_{bc}
$ with ${C^a}_{bc} \equiv G^{ad} C_{dbc}$ being the structure
coefficients of the zehnbein $d\theta^a = \frac12 {C^a}_{bc}
\theta^b \!\wedge\! \theta^c$. (Note the change in sign convention
here compared to above.) The frame derivative is $\partial_a
\equiv {E^i}_a (x) \partial_i$ (with $ {E^a}_i {E^i}_b =
\delta^a_b$). To determine the solution at any {\it given} spatial
point $x$ requires knowledge of an infinite tower of spatial
gradients; one should thus augment \Ref{EOM} by evolution
equations for $\partial_a G_{bc}, \partial_a {{\mathcal
F}}_{0bcd}, \partial_a {{\mathcal F}}_{bcde}$, etc., which in turn
would involve higher and higher spatial gradients.

The main result of concern here is the following: there exists a
{\it map} between geometrical quantities constructed at a given
spatial point $x$ from the supergravity fields $G_{\mu\nu}(x^0,x)$
and $\cA_{\mu\nu\rho}(x^0,x)$ and the one-parameter-dependent
quantities $g_{ab}(t), A_{abc} (t), \dots$ entering the coset
Lagrangian \Ref{Lag}, under which the supergravity equations of
motion \Ref{EOM} become {\it equivalent, up to 30th order in
height}, to the Euler-Lagrange equations of \Ref{Lag}. In the
gauge \Ref{Gauge} this map (or ``dictionary'') is defined by $t = x^0  \equiv \int dT/
\sqrt{G}$ and
\begin{eqnarray}\label{map}
g_{ab}(t) &=& G_{ab} (t,x) \,,\nonumber\\[3mm]
DA_{a_1a_2a_3}(t)  &=& {{\mathcal F}}_{0a_1 a_2 a_3} (t,x)\,,
\non[3mm] DA^{a_1 \dots a_6} (t) &=&   - \ft1{4!}
\varepsilon^{a_1\dots a_6 b_1 b_2 b_3 b_4} {{\mathcal F}}_{b_1 b_2
b_3 b_4} (t,x) \,, \non[3mm] DA^{b|a_1 \dots a_8 } (t)&=& \ft32
\varepsilon^{a_1\dots a_8 b_1 b_2} \big( {C^b}_{b_1 b_2} (x) +
\ft29 \delta^b_{[b_1}  {C^c}_{b_2] c} (x) \big)\,.
\end{eqnarray}
\subsection{Conclusions}
We have reviewed the finding that
the general solution of many physically relevant (bosonic) Einstein-matter
systems, in the vicinity of a space-like
singularity, exhibits a remarkable mixture of chaos and symmetry. Near the singularity, the behavior
of the general solution is  describable, at each (generic)
spatial point, as a billiard motion in an auxiliary Lorentzian space or, 
after a suitable ``radial''  projection, as a
billiard motion on hyperbolic space. This motion appears to be chaotic in many physically
interesting cases involving pure Einstein gravity in
any space-time dimension $D\leq 10$ and the particular Einstein-matter systems
arising in string theory. Also, for
these cases, the billiard tables can be identified with the Weyl chambers of
some Lorentzian Kac--Moody algebras. In the case of the bosonic sector of
supergravity in 11-dimensional space-time the underlying Lorentzian algebra
is that of the hyperbolic Kac--Moody group $E_{10}$, and there exists some
evidence of a correspondence between the general solution of the
Einstein-three-form system and a null geodesic in the infinite-dimensional
coset space $E_{10} / K (E_{10})$, where $K (E_{10})$ is the maximal
compact subgroup of $E_{10}$.

\vglue 8mm
\noindent{\bf Acknowledgement}

\noindent It is a pleasure to thank Sophie de Buyl and Christiane Schomblond for
their essential help in preparing these notes.

\appendix
\section{Kac-Moody algebras}

A KM algebra $\mathcal{G}(A)$ can be constructed out of a
generalized Cartan matrix $A$, (i.e. an $r\times r$ matrix such that $A_{ii} = 2, i=1
,...,r$, ii) $-A_{ij} \in \mathbb{N}$ for $i\neq j$
and iii)  $A_{ij} = 0$ implies $A_{ji} = 0$) according to the following rules for the
Chevalley generators $\{ h_i, e_i, f_i \}, {i=1,...,r}$: 
\begin{eqnarray}
\ [ e_i ,f_j ] &=& \d_{ij} h_i \nonumber \\
\ [h_i ,e_j] &=& A_{ij} e_j \nonumber \\
\ [h_i ,f_j] &=& -A_{ij} f_j \nonumber \\
\ [h_i ,h_j] &=& 0. \nonumber
\end{eqnarray}
The generators must also obey the Serre's relations, namely
\begin{eqnarray}
(\mathrm{ad}\,e_i)^{1-A_{ij}} e_j &=& 0 \nonumber \\
(\mathrm{ad}\,f_i)^{1-A_{ij}} f_j &=& 0 \nonumber
\end{eqnarray} and the Jacobi identity. $\mathcal{G}(A)$ admits a triangular decomposition
\be \mathcal{G}(A) = n_- \oplus h \oplus n_+   \ee where
$n_-$ is generated by the multicommutators of the form
$[f_{i_1},[f_{i_2},...]]$, $n_+$ by the multicommutators of the
form  $[e_{i_1},[e_{i_2},...]]$ and $h$ is the Cartan
subalgebra.\newline The algebras $\mathcal{G(A)}$
build on a symmetrizable Cartan matrix $A$ have been classified according to properties of their eigenvalues
\begin{itemize}
\item if $A$ is positive definite, $\mathcal{G(A)}$ is a finite
dimensional Lie algebra;
\item if $A$ admits one null eigenvalue and the others are all strictly positive, $\mathcal{G(A)}$ is an Affine KM algebra;
\item if $A$ admits one negative eigenvalue and all the others
are strictly positive, $\mathcal{G(A)}$ is a Lorentzian KM algebra.
\end{itemize}
A KM algebra such that the deletion of one node from its
Dynkin diagram gives a sum of finite or affine algebras is called an
\textit{hyperbolic} KM algebra. These algebras are all known; in particular,  there exists  no hyperbolic algebra with rang higher than 10. \newline

\end{document}